\documentclass[sigconf,accepted]{acmart}

\AtBeginDocument{%
  \providecommand\BibTeX{{%
    \normalfont B\kern-0.5em{\scshape i\kern-0.25em b}\kern-0.8em\TeX}}}

\setcopyright{none}
\copyrightyear{2019}
\acmYear{2019}

\acmConference[CVMP 2019]{the 16th ACM SIGGRAPH European Conference on Visual Media Production} {Dec.\ 17--18}{London, UK}

\begin{document}

\title[Automated Composition of Music for Movies]{Automated Composition of Picture-Synched \\ Music Soundtracks for Movies}

\author{Vansh Dassani}
\affiliation{%
  \institution{The University of Bristol }
  \city{Bristol}
    \postcode{BS8  1UB}
  \country{UK}
  }
\email{vansh.dassani.2015@my.bristol.ac.uk}

\author{Jon Bird}
\orcid{0000-0002-1681-1532}
\affiliation{%
  \institution{The University of Bristol }
  \city{Bristol}
    \postcode{BS8  1UB}
  \country{UK}
  }
\email{jon.bird@bristol.ac.uk}

\author{Dave Cliff}
\orcid{1234-5678-9012}
\affiliation{%
  \institution{The University of Bristol }
  \city{Bristol}
    \postcode{BS8  1UB}
  \country{UK}
  }
\email{csdtc@bristol.ac.uk}


\begin{abstract}

We describe the implementation of and early results from a system that
 automatically composes picture-synched musical soundtracks for videos and movies. We use the 
 phrase {\em picture-synched}\/ to mean that the structure of the automatically composed music is 
 determined by visual events in the input movie, i.e. the final music is synchronised to visual events and features such as cut transitions or within-shot key-frame events. Our system combines automated video analysis and computer-generated music-composition techniques to create unique soundtracks in response to the video input, 
 and can be thought of as an initial step in creating a computerised replacement for a human composer 
 writing music to fit the picture-locked edit of a movie. Working only from the video information in the movie, key features are extracted from the input video, using video analysis techniques, which are then fed into a 
machine-learning-based music generation tool, to compose a piece of music
from scratch. The resulting soundtrack is tied to video features, such as scene
transition markers and scene-level energy values, and is unique to the input video. Although the system we describe here is only a preliminary proof-of-concept, user evaluations of the output of the system have been positive. 

\end{abstract}




\maketitle

\section{Introduction}

Imagine that you have edited a 15-minute movie to the picture-lock stage, and now you need to source a continuous music soundtrack to be added in the final stages of postproduction: what are your options? You could attempt to find appropriate music on a stock-audio site such as MusicBed \cite{MusicBed}, but such sites tend to offer a large number of tracks that are five-minutes or less and very few that are over 10 minutes. Even if you do find a pre-recorded 15-minute music track to your liking, the chances are that you'd need to go back into edit to adjust the cut, so that key events in the movie fit well to the timeline imposed by the choice of music: but what if you don't want to do that (remember, your movie is already at picture-lock)? If you're a gifted {\em auteur}, you could compose your own soundtrack, fitting the music to the video, but that will likely take some time; and maybe you just don't have the music skills. If instead you have money, you could pay a composer to write a specifically commissioned soundtrack, synchronised to the cut of the video. But what if none of these options are practicable? In this paper we sketch a solution for your woes: we have created a proof-of-concept automated system that analyses the visual content of a video and uses the results to compose a musical soundtrack that fits well to the sequence of events in the video, which we refer to as a {\em picture-synched}\/ soundtrack. We have conducted a series of user studies, and the results demonstrate that our approach has merit and is worthy of further exploration. Because it's easier to refer to such a system via a proper noun, we have named our video-driven automated music composition system {\em Barrington}.\footnote{In honour of Barrington Pheloung (1954-2019), composer of clever themes, incidental music, and soundtracks for film and TV.}

Another way of framing the task performed by Barrington is as follows: in the production of promo videos for music singles, especially in genres such as disco, house, techno, and other forms of electronic dance music with a strongly repetitive beat structure, it is quite common to see videos edited in such a way that they're {\em cut to the beat}, i.e.\ where abrupt (hard-cut or jump-cut) transitions between scenes in the video are synchronised with the beat-pattern of the music. In those kind of promo videos, the pattern of switching between scenes within the movie is determined by the structure of the music: the music comes first, and the video is then cut to fit to the music. Barrington is a provisional system for automatically generating answers to the question posed by inverting this process: starting with a video, what music can be created that fits the style and structure of that video? Obviously, for any one video, there is an essentially infinite number of music tracks that could potentially be used as the soundtrack for that video, and deciding which ones are good, and which is best, is a subjective process requiring artistic judgement. Barrington incorporates a number of heuristics which are intended to improve the quality of its output. In its current proof-of-concept form, Barrington is certainly not offered as something that is likely to put professional music composers out of work anytime soon. More realistically, we think it likely that a later evolution of Barrington could conceivably be used to generate bespoke/unique soundtracks as a service for consumers, and an even more mature version could plausibly be used in creating targeted advertising. For example, the same video-advert could be shown to Customer A with a rock-music picture-synched soundtrack, and to Customer B with a classical-music picture-synched soundtrack, with both soundtracks produced by Barrington on the basis of prior knowledge of the customers' respective tastes in music. 

In Section \ref{sec:bckgnd} we review relevant background material. The technical notes on the design and implementation of Barrington are given in Section  \ref{sec:implement}. Then in Section \ref{sec:trials} we give details of our user studies and discuss the results. We close the paper with a discussion of further work in Section \ref{sec:concl}.

\section{Background}
\label{sec:bckgnd}

Computers have long been used as tools by musicians to experiment and produce music with. From simple applications that allow you to organise short audio clips into playable sequences, to complex audio synthesis models that can recreate the sounds of musical instruments, or even create new ones, computers have transformed our approach to making music. Today, the majority of musicians use digital audio workstations (DAWs), such as Logic Pro \cite{LogicPro} and Reaper \cite{Reaper}, which provide a single platform through which to record, edit and process audio.

Pythagoras is attributed with discovering that numerical ratios could be used to describe basic intervals in music, such as octaves and fifths, that form harmonic series \cite[Chapter~2]{weiss2007music}. This observation paved the way for formalising musical concepts using mathematics. In the 18\textsuperscript{th} Century, dice-based musical games, such as Mozart's Dice, were an example of algorithmic composition, in which a table of rules and a dice were used to randomly select small, predetermined sections of music to form a new piece \cite{10.2307/734136}. Today, those same games can run on a phone or computer, with no dice required.

In the 19\textsuperscript{th} Century, Ada Lovelace hypothesized that Babbage's Analytical Engine might be capable of algorithmic composition \cite[Note~A]{menabrea1842sketch}: 
\begin{quote}
    \textit{"Again, it might act upon other things besides number, were objects found whose mutual fundamental relations could be expressed by those of the abstract science of operations, and which should be also susceptible of adaptations to the action of the operating notation and mechanism of the engine. Supposing, for instance, that the fundamental relations of pitched sounds in the science of harmony and of musical composition were susceptible of such expression and adaptations, the engine might compose elaborate and scientific pieces of music of any degree of complexity or extent."}
\end{quote}

The idea of algorithmic music composition has been explored since the 1950's, but for the most part the composition algorithms were developed manually. Rule-based systems, such as Ebcioglu's CHORAL \cite{ebciouglu1990expert}, allowed for automated composition based around a set of pre-programmed rules, often referred to as a grammar. Musical grammars describe the form in which notes can be grouped or ordered to create sequences. The use of a formal grammar allows a computer to execute the algorithm, without any human intervention or direction. 





Arguably such algorithmic music composition techniques are examples of artificial intelligence (AI). However, more recent breakthroughs in AI, specifically deep learning methods, have vastly improved the quality and performance of computer-generated music systems. Researchers can use vast amounts of data, in this case in the form of existing music, to essentially `teach' a computer to solve a problem like a human would. Here, instead of a programmer defining the musical grammar, the machine observes and processes the input data to define the grammar on its own: i.e., it is able to `learn'. Van Der Merwe and Schulze used Markov chains to model chord duration, chord progression and rhythmic progression in music \cite{5492670}. Markov chains model sequences of events as discrete states and the transition probabilities between states, and do not rely on knowledge of other states to make a transition. More sophisticated Markov models have been useful for tasks such as accompaniment or improvisation, as shown by Morris {\em et al.} \cite{morris2008exposing} in their work on human-computer interaction for musical compositions, and Sastry's 2011 thesis \cite{sastry2011n} exploring machine improvisation and composition of tabla sequences. New deep learning models, such as BachBot \cite{liang2016bachbot} are capable of producing complex music that can be virtually indistinguishable from their human-produced counterparts.

One of the first practical applications of this technology has been creating royalty-free music for video content creators. Video content is rapidly becoming the preferred medium through which to advertise and communicate over the internet. Cisco estimates that by 2022, online video content will make up 82\% of consumer internet traffic \cite{cisco2018cisco}. As demand for video content grows, the content creation industry is looking for novel methods to improve and streamline their processes. In recent years, tools such as Animoto \cite{Animoto} and Biteable \cite{Biteable} have allowed novices to produce professional quality videos in just a few clicks, by offloading the more time consuming and skill-based aspects of video editing to machines. These tools often provide a selection of royalty-free music for users to easily add a soundtrack to their productions. However, these compositions have a fixed style and structure, which leaves limited, if any, scope for altering the soundtracks to better support the visuals.

Advertisers have also turned to video content to better convey meaning and engage with customers. Drawing on relationships between auditory and visual emotional responses, it could become easier to create emotionally engaging content, thereby improving the effectiveness of advertising campaigns. In the future, it is possible to envision social media advertising where the genre or style of the soundtrack is tailored to individual viewers. Video hosting platforms, such as YouTube \cite{YouTube} and Vimeo \cite{Vimeo}, could evolve to support high levels of real-time customisation, allowing different viewers to have their own unique experience of the same video content.

Computer vision is another field that has benefited from advances in machine learning, utilising vast datasets to infer features relevant to the task at hand. Research in this area has enabled many interesting applications, from identifying and tracking moving objects in videos (e.g., \cite{Nam_2016_CVPR}), to detecting human emotions through facial expressions (e.g. \cite{4624313}). Recently, Chu and Roy \cite{DBLP:journals/corr/abs-1712-02896} described a novel method of learning to identify emotional arcs in movies using audio-visual sentiment analysis, which opens up new opportunities for deriving meaningful information from video input.

Computationally-generated video from audio input is a well-established domain of study (see e.g. Beane's 2007 thesis \cite{beane2007generating}). But if computers can generate visual imagery to support music, why not the other way round? As far as we can determine, very little research has been published in this area. There has been some work on extracting image features such as contour, colour and texture to generate music, as in Wu and Li's study of image-based music composition \cite{wu2008study}, and this field is often referred to as image {\em sonification}. As the name suggests, these methods work on still images and therefore could, in principle, be applied to video content.

The lack of published work on this topic  was one motivation for our work; another was that the digital video industry is valued at over \$135 billion in the USA alone\cite{Magisto}, and represents a huge potential market for a solution. The contribution of this paper is to describe the design and implementation of Barrington, a proof of concept system that our initial user evaluation indicates could, with further refinement, prove to be extremely time and cost effective for content creators working with prohibitive music licensing costs, or those without the technical expertise to create or edit a soundtrack for their video.

\section{Implementation}
\label{sec:implement}

\subsection{Design}
\label{sec:design}

 In traditional film-making, directors and editors work closely with composers to develop a soundtrack that fits with their narrative intent. Similarly, our intention was to create a tool that was not totally autonomous but rather could be guided by visual content creators.
 
 
 
 
 Barrington is primarily written in the Python 3 programming language, due to the availability of a wide range of open-source libraries; as is made clear later in this paper, some aspects of Barrington are controlled by shell-scripts consisting of operating-system commands.

\subsection{Scene Transition Detection}
Scene transitions are the most easily identifiable markers of progression in a video, and served as the starting point for the analysis. Scene transition detection, often also referred to as shot boundary detection, is a difficult task for computers, due to the variety of transition methods video editors have at their disposal. Abrupt transitions follow a predictable pattern, as the change in content from one frame to another is sudden and usually extensive. Gradual transitions often utilise visual effects to blend together two scenes into a smooth manner, such as cross-dissolves or fades, over any length of time, and with very little change in content from frame to frame. This variability makes it difficult to create a single method of detecting every type of transition. Transition detection algorithms typically look at pairs of frames, computing a score for each using some defined similarity measure, then evaluating these scores against some threshold to determine if a transition has occurred.


Following \cite{zhang1993automatic}, Barrington implements two basic scene detectors, one to detect abrupt transitions ({\em cut detection}) and the other to detect fades to/from black ({\em fade detection}). 

Cuts are detected by computing  the average change in hue, saturation and intensity values from one frame to the next. This aggregate value will be low between frames with little change in content, and high between frames with large changes. A threshold is specified such that frames above the threshold are marked as abrupt transitions. 

Fades are detected using RGB intensity. For each frame, the R, G, and B pixel values are summed to produce an intensity score. A threshold is specified such that the RGB intensity values in the majority of frames within a scene lie above said threshold; if the RGB intensity for a frame falls below the threshold, it is marked as the start of the fade, and the next frame to return above the threshold is marked as the end. This remarkably simple technique works effectively, as any fade to/from black will inevitably contain at least one black frame, with an RGB intensity value of 0.


PySceneDetect \cite{PySceneDetect} was the open-source library chosen to implement the scene detection algorithms described above. It provides methods for performing both fade detection and cut detection, each outputting a list of scenes identified with their start and end times. Barrington uses PySceneDetect's default threshold values of 12 and 30 for the fade and cut detection respectively, outputting a final list of scenes detected through both detection methods. The library is also able to export clips of individual scenes, which are used for scene-level analysis discussed later in Section \ref{sec:scene-energy}.


\subsection{Loop-based Soundtracks}

The list of scenes output from scene detection provide a starting point on which to base the temporal structure of the soundtrack. Taking a simplistic view, we can split the overall soundtrack into sections commonly found in popular music, such as the intro, verse, chorus and coda (outro), and map each scene to a section of music: the first scene being the intro and the last the coda. Mapping the rest of the scenes is more complicated, as different videos will have different numbers of scenes.

In order to gain an understanding of how best to approach this, we first created a simple rule-based system that loops some pre-existing samples of music, with the middle scene (or middle pair of scenes for cases where the total number of scenes is even) arbitrarily chosen as the point at which the soundtrack peaks, and every other scene transition either introduces or removes a particular sample. This is analogous to the way music is composed professionally, as repetition is often a key part of musical structure. These loop-based sequenced soundtracks were an important early step in developing Barrington because they allowed us to explore and demonstrate video-driven sequence-assembly, but they offer only a coarse-grained approach to music composition. 

To give Barrington more creative scope, we switched to using fully-fledged music generation systems. We initially used Magenta PolyphonyRNN, a recurrent neural network, based on the BachBot architecture described in \cite{liang2016bachbot}. It is able to compose and complete polyphonic musical sequences using a long short-term memory (LSTM) generative model, without having to explicitly encode musical theory concepts. BachBot's success is based on modelling melodies and harmonisation independently. It is able to learn a representation of harmonisation based on samples from a corpus, and apply it to any melody. In user tests it was capable of producing music virtually indistinguishable from Bach's compositions. Unfortunately, the computational costs of PolyphonyRNN meant it was not suitable for our system: run-times lasting a week or more could be consumed in training the system, and the results were often disappointing. We therefore switched to IBM's Watson Beat.

\subsection{Music Generation: Watson Beat}

IBM's Watson Beat \cite{IBMWatsonBeat} is designed to be a compositional aid for musicians, helping them generate new ideas they can then adapt to compose a piece of music, although it is capable of producing a complete piece on its own. Watson Beat provides more granular control over the composition compared to PolyphonyRNN; it composes parts for multiple instruments at once, and allows users to define the composition's mood, instrumentation, tempo, time signature and energy. This made it a much more versatile tool for developing Barrington.


A brief overview of the process of developing Watson Beat is available on IBM's website \cite{IBMWatsonBeat}, in which the developers describe using supervised learning on a vast amount of data relating to each composition's pitch, rhythm, chord progression and instrumentation, as well as labeled information on emotions and musical genres. A key feature of Watson Beat is its ability to arrange the musical structure in a manner that provides the user control over individual sections of music, though it is unclear if this was manually encoded or learned from training data.

Watson Beat requires two inputs to compose a piece of music: a MIDI file with a melody to base the composition on; and a \textit{.ini} file that specifies the structure of each section of the composition. The \textit{.ini} file for each composition must start with the following parameters:

\renewcommand{\baselinestretch}{1}\normalsize
\begin{quote}
\begin{itemize}
\item \textbf{Composition duration}: in seconds.
\item \textbf{Mood}: which is selected from one of the 8 pre-configured options.
\item \textbf{Complexity}: three options (simple, semi-complex, or complex) which determines the complexity of the chord progressions in the composition.
\end{itemize}
\end{quote}

This is followed by a list of parameters for each section of the composition, which are explained below:

\renewcommand{\baselinestretch}{1}\normalsize
\begin{quote}
\begin{itemize}
\item \textbf{Section ID}: a series of consecutive integers starting from 0 identifying the current section.
\item \textbf{Time signature}.
\item \textbf{Tempo}: in BPM.
\item \textbf{Energy}: three options (low, medium or high) which represents the number of active instruments for the current section. The number of instruments (also referred to as layers) each energy option uses can vary, and depends on the selected mood. For example, low could signify 1-3 instruments for one mood option, but 2-5 for another.
\item \textbf{Section duration}: in seconds (allows for a range to be specified, such as 10 to 20 seconds, for situations where a precise duration is not required).
\item \textbf{Direction}: set to either up or down, and determines whether instrument layers will be added or removed during the section.
\item \textbf{Slope}: one of three options (stay, gradual or steep) which control the rate of change of the Direction, that is, the rate at which layers are added or removed.
\end{itemize}
\end{quote}

Watson Beat only requires a duration range to generate each section of music, and calculates a best-fit phrase length, tempo and time signature, from a pre-specified range of options for the given mood, for durations within this range. This calculation is non-deterministic as it utilises a randomly generated number to pick from the computed values. Given that section durations for the autogenerated soundtracks must match exactly with the video, the best-fit calculation is not ideal as it can often lead to an incorrect section duration. 

To counter this, Barrington incorporates a Python function to compute the correct tempo and time signature required to produce the specified section duration. The function loops through every possible combination of parameters and selects those where the number of phrases is a whole number and the calculated duration equals the desired duration. These lists of valid combinations are generated for each scene, and then trimmed to only contain options that ensure a consistent tempo throughout the soundtrack. For each section, one of these possible combinations is randomly selected for use, ensuring variation in the resulting soundtracks. To generate the soundtrack, Barrington invokes a Bash shell script that runs Watson Beat, specifying the \textit{.ini} file, an existing MIDI file with an input melody, and the output folder for the generated MIDI files.

As described thus far, Barrington is able to automatically produce synchronised soundtracks with discernibly different sections, only requiring the \textit{energy}, \textit{direction} and \textit{slope} parameters to be manually input. These parameters, along with the composition's mood, play a role in the subjective interpretation of how the soundtrack supports the visual content, therefore it is important for users to have control over them. Nevertheless, for the system to operate at higher levels of autonomy, it must also be capable of determining these values itself, directly from the input video.

\subsection{Determining the Energy of a Scene}
\label{sec:scene-energy}
There are no definitive criteria to determine the energy of a scene of video or a section of music; the parameter is intrinsically dependant on the music generation tool being used. Watson Beat uses energy, direction and slope to define the number of instrument layers active in a given section of music, and how this number should vary through the composition, so energy values such as low, medium and high are only relevant in the context of a Watson Beat composition. Traditionally, tempo is used to convey energy - a high tempo is considered more energetic than a low tempo - so it stands to reason that any representation of energy derived through video analysis should control tempo, as well as the instrumentation parameters.

A naive approach to this problem could be to use activity or `busyness' within a scene to determine energy. For example, a scene with lots of objects can be considered more energetic than one with very few. The downside of naive approximations is that once again there is no consistent way of comparing the results across different videos. Without a standardised metric, the only method of determining if a scene has  low or high energy is to compare it with others within the same video.

To test the effectiveness of a naive object-counting approach, we wrote a Python function that uses ImageAI \cite{ImageAI}, an open-source computer vision library, and version 3 of the YOLO object detection model 
 \cite{DBLP:journals/corr/RedmonDGF15}. YOLO utilises a single convolutional neural network (CNN) for training and detection. which are often used for image classification tasks. In mathematics, a convolution expresses how the shape of some function is affected as it is shifted across a \textit{static}, or non-moving, function, i.e. `blending' the two together. CNNs treat the input image as this static function and pass lots of different functions over it, each of these being a \textit{filter} that looks for a specific feature in the image. Each filter is passed over small sections of the image at a time, to build up a \textit{feature map} that tracks the locations at which various features occur across the image. 
 
YOLO is able to learn a general representation for different object classes, while optimising detection performance during training, resulting in faster, more precise detection. YOLO subdivides an image into a grid, using features from across the whole image to predict the bounding box for every object class, and a confidence score that encodes the probability of that box containing an object. The clips of individual scenes are processed one by one to produce a listing of the number of objects detected in each scene. Barrington then calculates the mean and standard deviation of the results to label the scenes low, medium or high energy: scenes with a number of objects fewer than one standard deviation below the mean are classified as low energy; scenes within the mean plus or minus one standard deviation as medium-energy; and scenes one standard deviation or greater above the mean as high-energy. A similar approach is used to match an associated tempo; the tempo range for a particular mood is split into three and matched the same way as the energy value, then a tempo within this range is selected from the list of valid tempo options. This approach does not work for all scenes, as it relies on a tempo within the sub-range to be valid for the particular scene duration, which may not be the case, but it is a useful first approximation. 

The Watson Beat \textit{direction} and \textit{slope} parameters are selected by comparing each scene's energy value to the next, using a heuristic decision tree, details of which are given in \emph{Anonymous Citation}. 

For the final scene in the video, the function assumes the next scene has the same energy value as the current. Note that if `stay' is selected for the Slope parameter, it doesn't matter what direction is selected as the number of instrument layers stays constant through the section.

\subsection{Post-Processing}

The final step involves post-processing the MIDI files and exporting a video with the generated soundtrack. Watson Beat outputs a separate MIDI file for each instrument within each section of music, allowing composers to rearrange the piece and select the instruments to use when converting to an audio file. While this option is still available for more technical users, the Barrington default post-processing step converts these MIDI files into a single audio file containing the final soundtrack, as well as producing a copy of the input video with the new soundtrack attached.

The MIDI specification allows for instruments to be declared in a message at the start of a file, however the instrument labels used by Watson Beat do not map directly to instruments listed in the General MIDI 1 (GM1)\footnote{\url{https://www.midi.org/specifications-old/item/gm-level-1-sound-set}} sound set. Barrington takes all the generated MIDI files as an input, assigning the correct tempo to each section using the reference structure, and the correct instrument using a simple mapping based on recommendations found in Watson Beat's documentation (\url{https://github.com/cognitive-catalyst/watson-beat/blob/master/customize_ini.md}). It then concatenates them into a single file and uses FluidSynth \cite{FluidSynth}, an open-source software synthesizer, to play and record the soundtrack to audio. A Bash script exports the final video, using ffmpeg  \cite{ffmpeg}, a cross-platform video conversion tool, to attach the soundtrack to the input video. Samples of output from Barrington are available online at \emph{Anonymous Website}.

\section{User Evaluation}
\label{sec:trials}

In this section we describe the user studies we carried out to evaluate the current version of Barrington. We begin by describing the methodology used, and the goal of the studies, after which we present and discuss the results. We conclude with reflections on the project thus far and its outcomes.

\subsection{Methodology}

Ultimately, a system like Barrington is only useful if the end user enjoys its outputs, so this is a primary measure by which to judge success. Our goal was to determine how the Barrington-generated soundtracks affected a user's experience of a video. There are few, if any, objective methods for measuring this, therefore a user study was carried out to measure subjective responses to a selection of Barrington's outputs, adapting the procedure used by Ghinea and Ademoye \cite{ghinea2012sweet}. They used a questionnaire to measure a user's quality of experience (QoE) of visual media when they were presented alongside stimuli in other sensory modalilties, such as olfaction. The QoE study they performed was particularly well suited for evaluating Barrington, as it also investigated the role of synchronisation of the different sensory stimuli on users' QoE, that is, how temporal offsets between the sensory stimuli affected users' reported QoE. A key feature of Barrington is the synchronisation of changes in auditory and visual content, so it is important to assess  whether users were able to detect this synchronisation and, if so, how it affected their QoE. 

Another element of Barrington that we wanted to evaluate is the relationship between the \emph{energy} parameter of a section of the soundtrack, and the corresponding scene's visual content. In particular, we wanted to test whether the object detection method discussed in Section \ref{sec:scene-energy} is a suitable proxy for the mood or intensity of a scene. 

Finally, it is important to assess how Barrington's output compares with existing solutions in the form of royalty-free music that is not picture-synched. We wanted to test whether the soundtracks generated by Barrington produced greater QoE than royalty-free music whose structure is not synched with the video content.

How these key feautures affected users' QoE was tested using an approach inspired by a Dual-Stimulus Impairment Scale (DSIS), a standard video testing methodology recommended by the International Telecommunications Union (ITU) \cite{series2012methodology}. In a DSIS test the viewer sees a reference video and a video with an \textit{impairment} or alteration, and is asked to indicate on a Likert scale how irritating they find this impairment. Instead of only measuring a user's experience of the impaired video, we asked participants to rate both the reference and the impaired videos, in order to determine whether their QoE was affected. Participants were not informed which video was which, to prevent any bias when responding to the questions.

\subsubsection{Hypotheses}

The user evaluation was carried out to test two hypotheses. 

First, we wanted to investigate the effect of synching a soundtrack with video scene changes, hypothesising that:\\
\emph{H1} -- users will report higher QoE for videos combined with synchronised soundtracks generated by Barrington than for videos combined  with unsynchronised soundtracks. 

Second, we sought to investigate whether the method of selecting the soundtrack's energy parameters was appropriate, hypothesising that:\\
\emph{H2} -- users will perceive a correlation between the auditory and visual content of the synchronised soundtracks generated by Barrington.

\subsubsection{Participants}

The 26 participants who volunteered to take part in this study consisted of 17 males and 9 females between the ages of 20 and 24. All the participants were either undergraduate or graduate students at \emph {Anonymous University}, the majority of whom study in the Computer Science Department. Each participant was provided with information and task instruction sheets explaining  the goal of the project (which can be viewed in \emph{Anonymous Citation}) before taking part, and was left alone in a private meeting room to proceed through the experiment at their own pace. 

\subsubsection{Procedure}
Each participant completed four tasks, the first three focusing on, respectively: picture-synchronisation, the relationship between auditory and visual energy, and a comparison with royalty-free soundtracks. The final task consisted of a short six item questionnaire relating to videos and soundtracks. 


For the first three tasks, each participant was presented with two videos, the reference and the impaired, which had identical visual content but different soundtracks. They were asked to watch each video in turn and complete the seven-item questionnaire shown below. Each item on the questionnaire required participants to select one option from a five-point Likert scale. For statements 1-6, participants were asked to indicate their level of agreement (LoA) by choosing one of the following options: \textit{Strongly agree}, \textit{Agree}, \textit{Neither agree nor disagree}, \textit{Disagree}, or \textit{Strongly disagree}. Statement 7 required participants to rate the synchronisation of the soundtrack and the scene transitions by selecting one of the following options: \textit{Too early}, \textit{Early}, \textit{At an appropriate time}, \textit{Late}, or \textit{Too late}.

\begin{quote}
\begin{enumerate}
\item \textit{The soundtrack added to the video experience}\label{stm:1}
\item \textit{The soundtrack was relevant to the video}\label{stm:2}
\item \textit{The soundtrack was enjoyable}\label{stm:3}
\item \textit{There were noticeable points in the soundtrack where the melody/rhythm changed}\label{stm:4}
\item \textit{The mood/intensity of the soundtrack reflected the mood/intensity of the video content}\label{stm:5}
\item \textit{The synchronisation between the audio and video improved the overall experience}\label{stm:6}
\item \textit{In relation to the scene transitions in the video, the melody/rhythm of the music changed}\label{stm:7}
\end{enumerate}
\end{quote}

In the final task, participants were presented with the following six statements and asked to indicate their LoA on a five-point Likert scale: \textit{Strongly agree}, \textit{Agree}, \textit{Neither agree nor disagree}, \textit{Disagree}, or \textit{Strongly disagree}.

\begin{quote}
\begin{enumerate}
\item \textit{Soundtracks are an important aspect of video content}\label{stm:4-1}
\item \textit{Soundtracks improve the experience of watching video content}\label{stm:4-2}
\item \textit{The mood of a soundtrack should match the mood of the video content}\label{stm:4-3}
\item \textit{A bad soundtrack negatively affects the viewing experience}\label{stm:4-4}
\item \textit{There was an obvious relationship between the soundtracks and the video content in the previous examples}\label{stm:4-5}
\item \textit{The synchronisation between the soundtrack and video improved the overall experience}\label{stm:4-6}
\end{enumerate}
\end{quote}

\subsubsection{Test Videos}
We created three 60-second duration, soundtrack-free videos for this study - one for each of the first three tasks. The videos consisted predominantly of open source wildlife and nature clips, chosen because they best complemented the \textit{Inspire} mood that had been selected in Watston Beat and which generates calming, cinematic music. The only exceptions to this were a clip of a busy pedestrian crossing and a clip of a herd of wildebeest, which were used because they were a sharp contrast to the relatively motion-free nature scenes, and specifically chosen to output a high energy parameter when object detection was used. Each soundtrack-free video was processed by Barrington to create the soundtracks for the reference videos, while the process for creating each of the impaired videos is described below.

{\bf Task 1} investigated the effect of synchronisation of the video scene changes and the soundtrack on users' QoE and tested H1. The video used contained three scene transitions, which provided ample opportunity for the participants to determine if the audio and video content were synchronised. The reference video the soundtrack was synchronised with the video scene changes. In the impaired video, we altered the scene-transition timings so the soundtrack transitions were three seconds earlier than their corresponding scene transitions, then used the system to generate a new soundtrack based on the alteration.

{\bf Task 2}  investigated the effect of the value of the energy parameter and investigated H2 The video contains just two scenes, one with  low energy and the other high. The high energy scene contains lots of moving objects, which intuitively corresponds to an increased scene intensity. To create the impaired video the energy parameters were swapped from high to low and  {\em vice versa}. (N.B. abrupt scene transitions were ignored in this video, due to the fact that one of the clips used consisted of multiple hard cuts within seconds of each other, which led to very short sections of music that clashed quite heavily and sounded bad.)

{\bf Task 3}  compares a video with a Barrington generated soundtrack to a video with a royalty-free soundtrack and also investigated H1. For the impaired video, a song was selected from Bensound \cite{Bensound}, an online royalty-free music provider, and trimmed to the duration of the video.



\subsection{Results}

\subsubsection{Task 1: Synchronisation}

Two aspects of synchronisation were tested: a) the detection of auditory and visual transitions; and b) the effect of their synchronisation on the participants' enjoyment of the video. For the reference video in Task 1, 26/26 (100\%) participants responded that the music changed, or progressed, at an appropriate time in relation to scene transitions, while only 3/26 (12\%) believed the same for the impaired video, where the soundtrack transitioned three seconds prior to the visual scence transitions.

\begin{figure}[h!]
\centering
\includegraphics[width=0.4\textwidth]{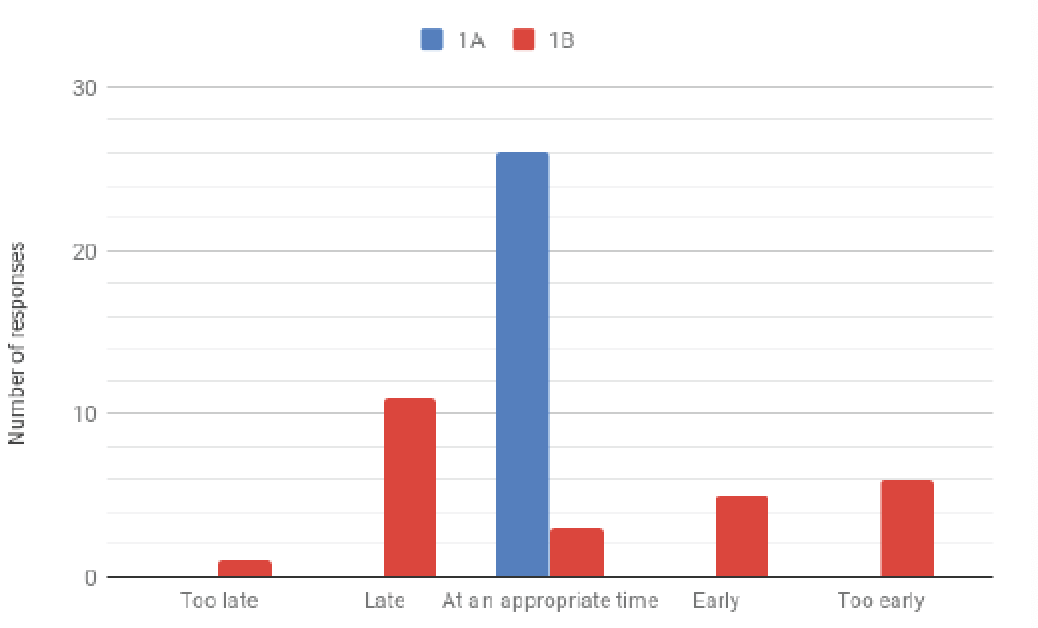}
\caption{Participant responses to the statement \textit{In relation to the scene transitions in the video, the melody/rhythm of the music changed}, in Task 1 of the user study.}
\label{res:q1-7}
\end{figure}

This initially suggests that participants were able to recognise the points at which musical transitions occurred, and determine whether they were in-sync with visual scene transitions. However, it is clear from the results in Figure \ref{res:q1-7} that while participants were able to detect that the audio and visuals were not synchronised, they were poor at judging whether the music transition occurred before or after the scene transition. 46\% of participants responded with \emph{late} or \emph{too late} to question \ref{stm:7} even though the soundtrack was configured to transition three seconds earlier than the video.

\begin{figure}[h!]
\centering
\includegraphics[width=0.4\textwidth]{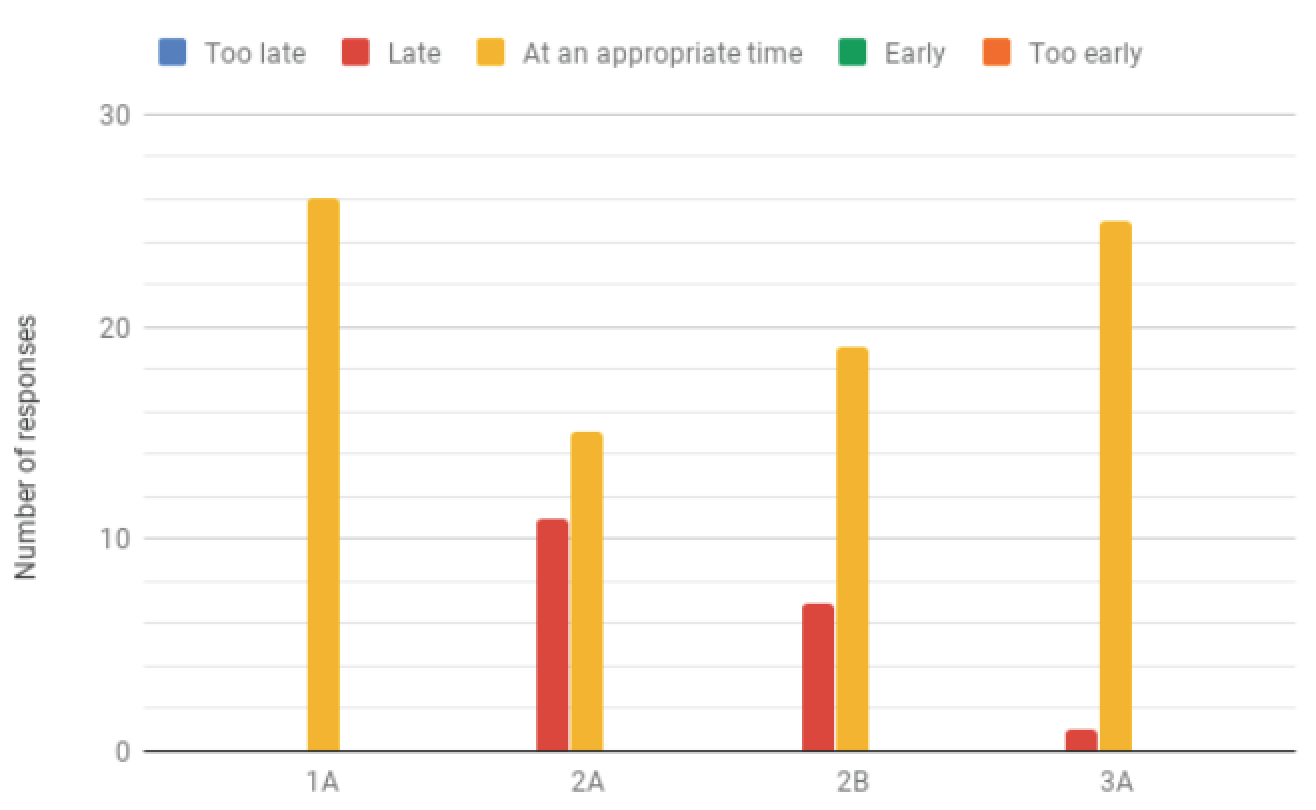}
\caption{Participant responses to the statement \textit{In relation to the scene transitions in the video, the melody/rhythm of the music changed}, across all videos produced via Barrington.}
\label{res:sync_all}
\end{figure}

When looking at the response to this question across all videos synchronised by Barrington,
Figure \ref{res:sync_all} shows that the majority of participants thought the synchronisation was appropriately timed.

It's interesting to note the remaining participants all selected \emph{late}, suggesting that participants may differ from one another in what they consider as being appropriately synchronised. Additionally, the videos for Task~2 elicited different timing responses, which is unexpected because the timing parameters for both were identical. 

This is possibly a result of the musical transitions being too gradual, leaving participants unclear as to when a transition occurred. Or it is perhaps an indication that the end of a scene is not the best marker for a musical transition, although it is not immediately clear what an alternative may be.

\begin{figure}[h!]
\centering
\includegraphics[width=0.4\textwidth]{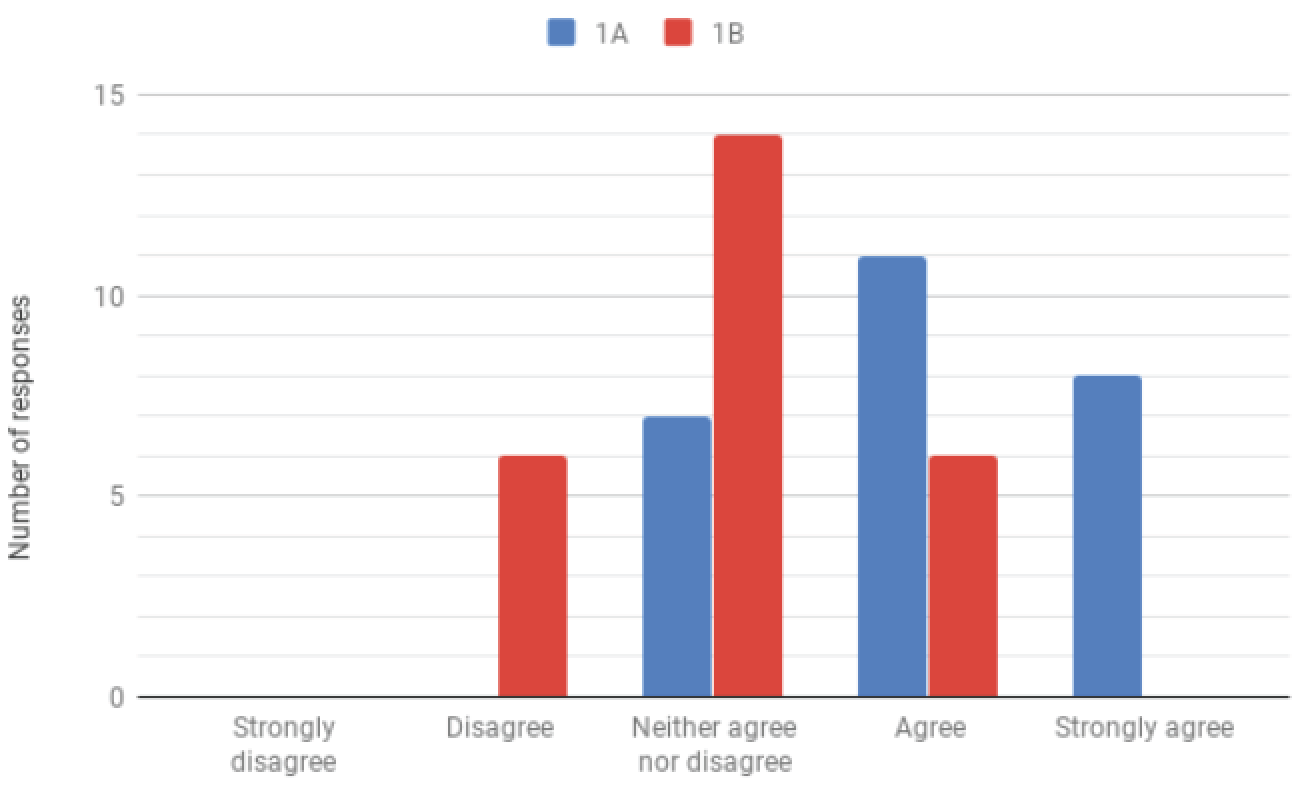}
\caption{Participant responses to the statement \textit{The synchronisation between the audio and video improved the overall experience}, in Task 1 of the user study.}
\label{res:q1-6}
\end{figure}

Figure \ref{res:q1-6} presents participant responses to statement \ref{stm:6}, which ascertains whether the soundtrack has enhanced their enjoyment of the video. The impaired video, 1B, elicited a predominantly neutral response, which could suggest either that the delay between musical and visual transitions is not significant enough to adversely affect enjoyment, or that a lack of synchronisation does not adversely affect enjoyment. This makes intuitive sense, as one would assume a large proportion of videos on media platforms contain unsynchronised soundtracks. In contrast, the reference video 1A received predominantly positive responses, suggesting that synchronisation can enhance the quality of experience.

\begin{figure}[h!]
\centering
\includegraphics[width=0.4\textwidth]{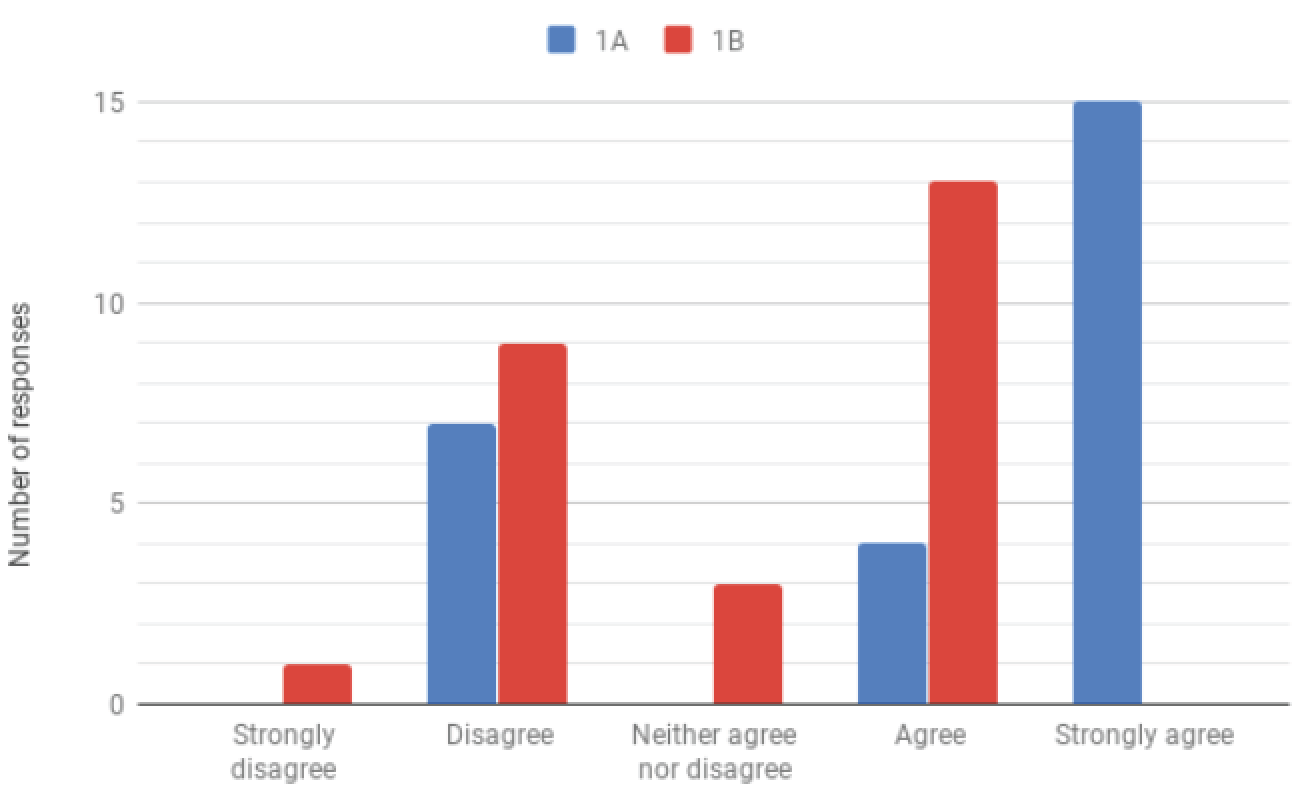}
\caption{Participant responses to the statement \textit{There were noticeable points in the soundtrack where the melody/rhythm changed}, in Task 1 of the user study.}
\label{res:q1-4}
\end{figure}

Looking at participant responses to statement \ref{stm:4}, plotted in Figure \ref{res:q1-4}, can give us further insight into why the above results neither supports nor disproves the hypothesis that synchronisation can enhance the experience.

Only 50\% of participants were able to detect musical transitions in the impaired video 1B, compared with 73\% for the reference video 1A. This discrepancy in responses is unexpected, as intuition suggests participants would respond the same way for 1A and 1B, given that both soundtracks contain three transitions. 

It is possible this is due to the different sections within those specific soundtracks sounding very similar to one another, however it is not possible to determine this from the results of this study. Ambiguity regarding participants' interpretations of the word \textit{changed}\/ could also have played a role, however none of the participants brought this up in the informal post-study feedback conversations. It seems that participants found it difficult to detect melody and rhythm, despite overall responding with a more positive perceived experience for picture-synched videos.

\subsubsection{Task 2: Energy and Mood/Intensity}

\begin{figure}[h!]
\centering
\includegraphics[width=0.4\textwidth]{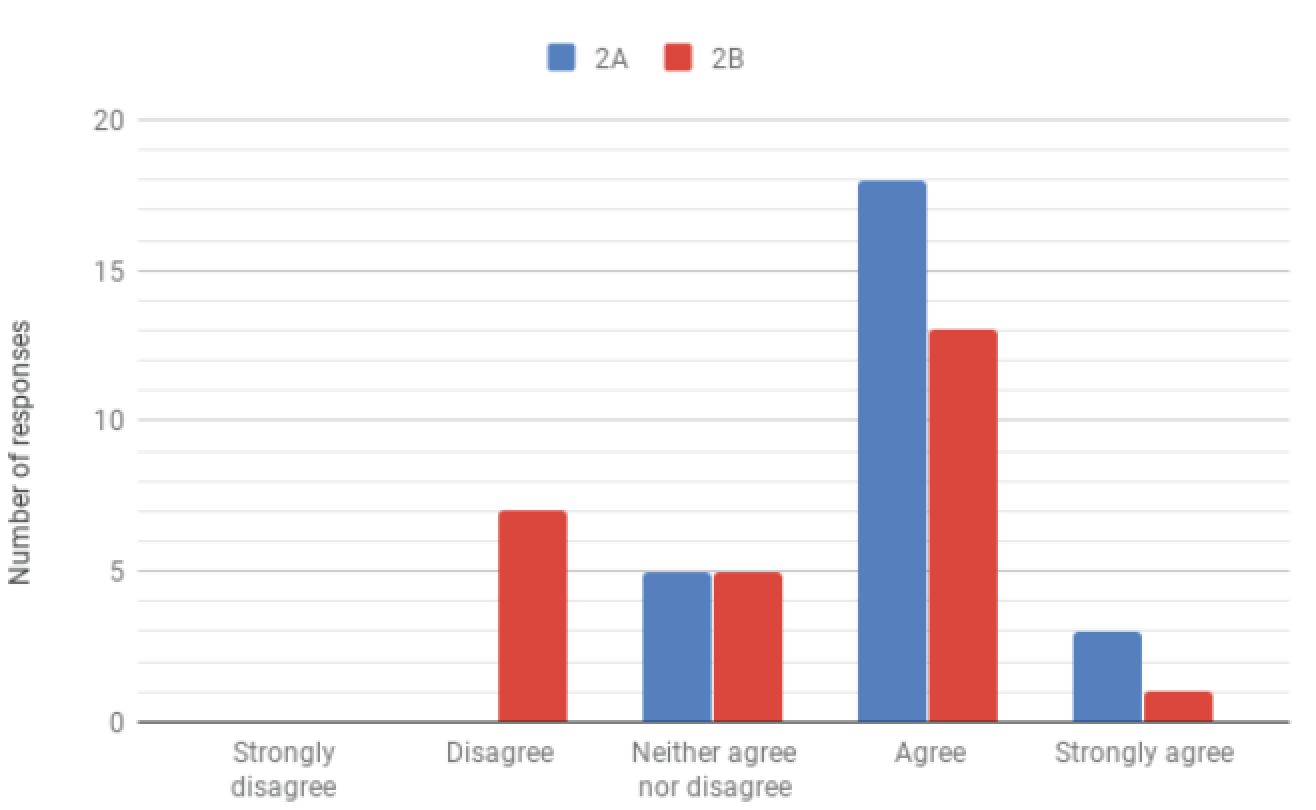}
\caption{Participant responses to the statement \textit{The mood/intensity of the soundtrack reflected the mood/intensity of the video content}, in Task 2 of the user study.}
\label{res:q2-5}
\end{figure}

This test was to determine whether the number of objects in a scene is a good proxy for how the soundtrack's energy levels should reflect the visual intensity of a scene. Figure \ref{res:q2-5} shows the results for the reference and impaired videos in Task 2 of the study. Only a small proportion (27\%) of participants provided the expected response disagreeing with statement \ref{stm:5} for the impaired video, whereas 81\% agreed for 1A and 54\% agreed for 1B. The results imply that the energy parameter has a minimal effect on the perceived mood/intensity of visual content, although it is promising to note that some participants did perceive a difference when the change was made, suggesting a more refined method could produce better results.

This raises the question of whether a soundtrack should reflect the mood/intensity of visual content: Section \ref{sec:task4} provides some insight on this.

\subsubsection{Task 3: Comparison with Royalty-free Music}

\begin{figure}[h!]
\centering
\includegraphics[width=0.4\textwidth]{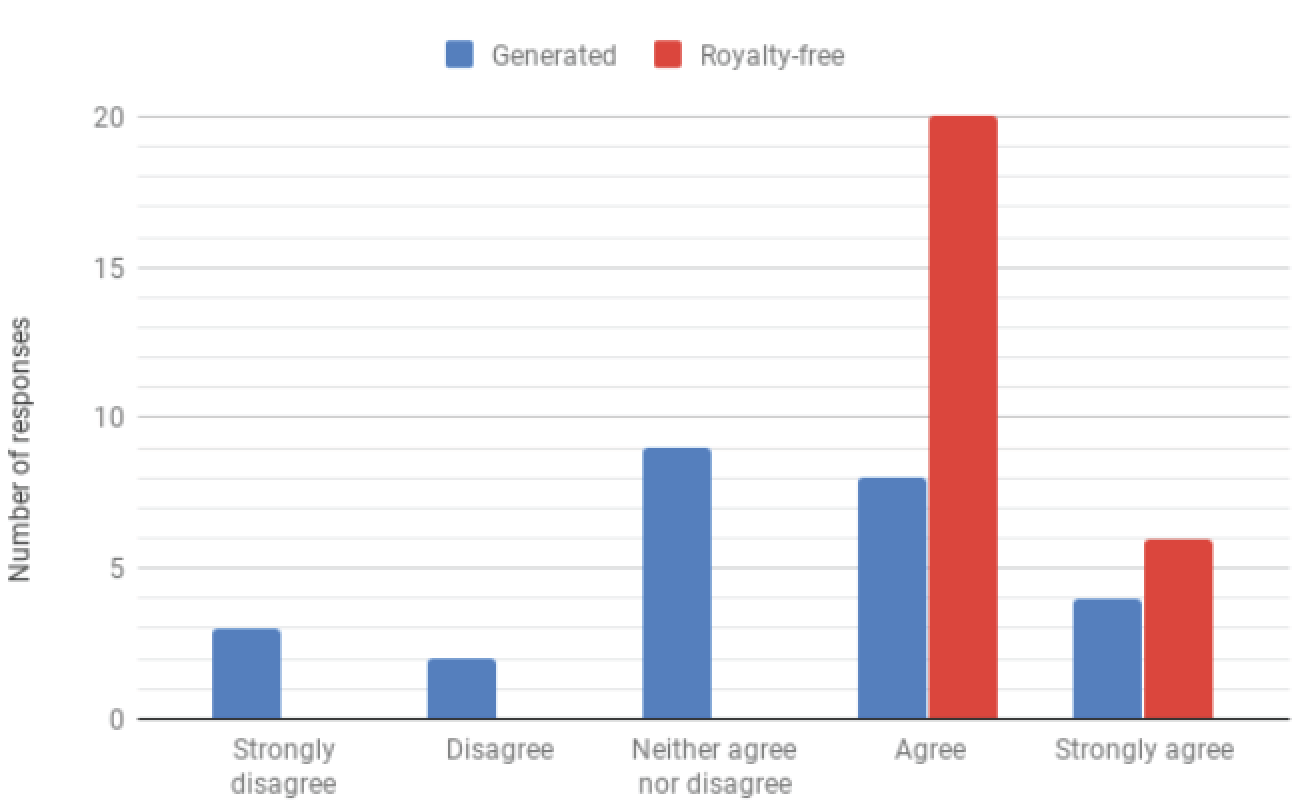}
\caption{Participant responses to the statement \textit{The mood/intensity of the soundtrack reflected the mood/intensity of the video content}, in Task 3 of the user study.}
\label{res:q3-5}
\end{figure}

In Task 3, participants responded more positively towards the royalty-free soundtrack on statements relating to the mood and relevance of the soundtrack with respect to the visual content. Figure \ref{res:q3-5} shows that 100\% of participants agreed that the mood/intensity of the royalty-free soundtrack reflected the mood of the visual content, while only 46\% thought the same for the Barrington-generated soundtrack.

\begin{figure}[h!]
\centering
\includegraphics[width=0.4\textwidth]{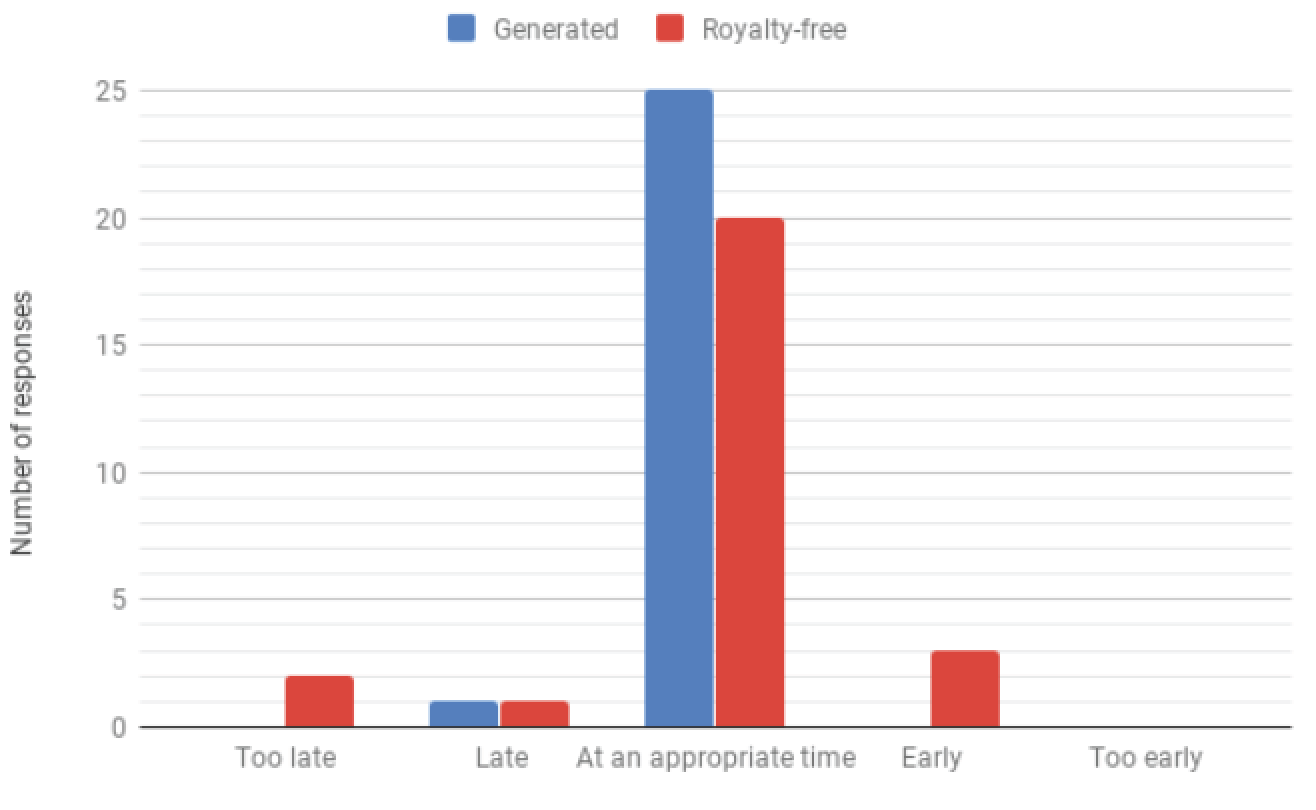}
\caption{Participant responses to the statement \textit{In relation to the scene transitions in the video, the melody/rhythm of the music changed}, in Task 3 of the user study.}
\label{res:q3-7}
\end{figure}

\begin{figure}[h!]
\centering
\includegraphics[width=0.4\textwidth]{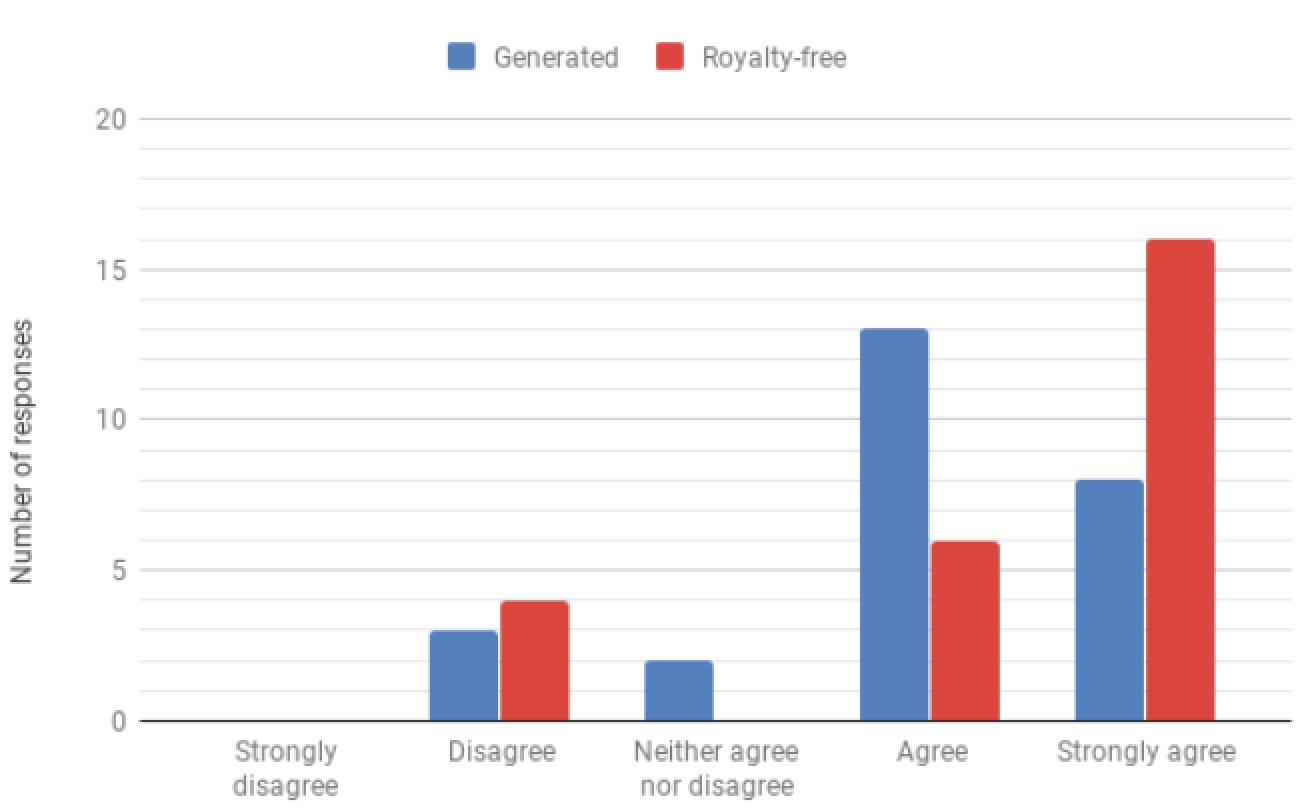}
\caption{Participant responses to the statement \textit{The soundtrack was enjoyable}, in Task 3 of the user study.}
\label{res:q3-3}
\end{figure}

This result reinforces the prior suspicion that the object-detection proxy for the soundtrack's energy level captures a limited representation of the mood of a scene in a video. However, it could also be a reflection of the audio quality of the music produced by the system. The royalty-free soundtrack was professionally composed and instrumented, whereas the generated soundtracks use a basic software synthesizer to apply instrument sounds as a post-process, a limitation imposed by Watson Beat. When it comes to synchronisation, participants were in agreement that the timing of the generated soundtrack was more accurate than the royalty-free one (see Figure \ref{res:q3-7}), although this did not have the expected result on user enjoyment, which once again could have been due to the quality of the soundtrack. Figure \ref{res:q3-3} shows a negligible difference (4\%, or 1 participant) in the enjoyment of the generated vs royalty-free soundtrack. This study was not intended to compare the quality of generated soundtracks against professionally produced royalty-free samples, but it is encouraging to note that the Watson Beat soundtracks were no less enjoyable than their counterparts.

\subsubsection{Task 4: Participants' Reflection on Video Soundtracks}
\label{sec:task4}

\begin{figure}[h!]
\centering
\includegraphics[width=0.4\textwidth]{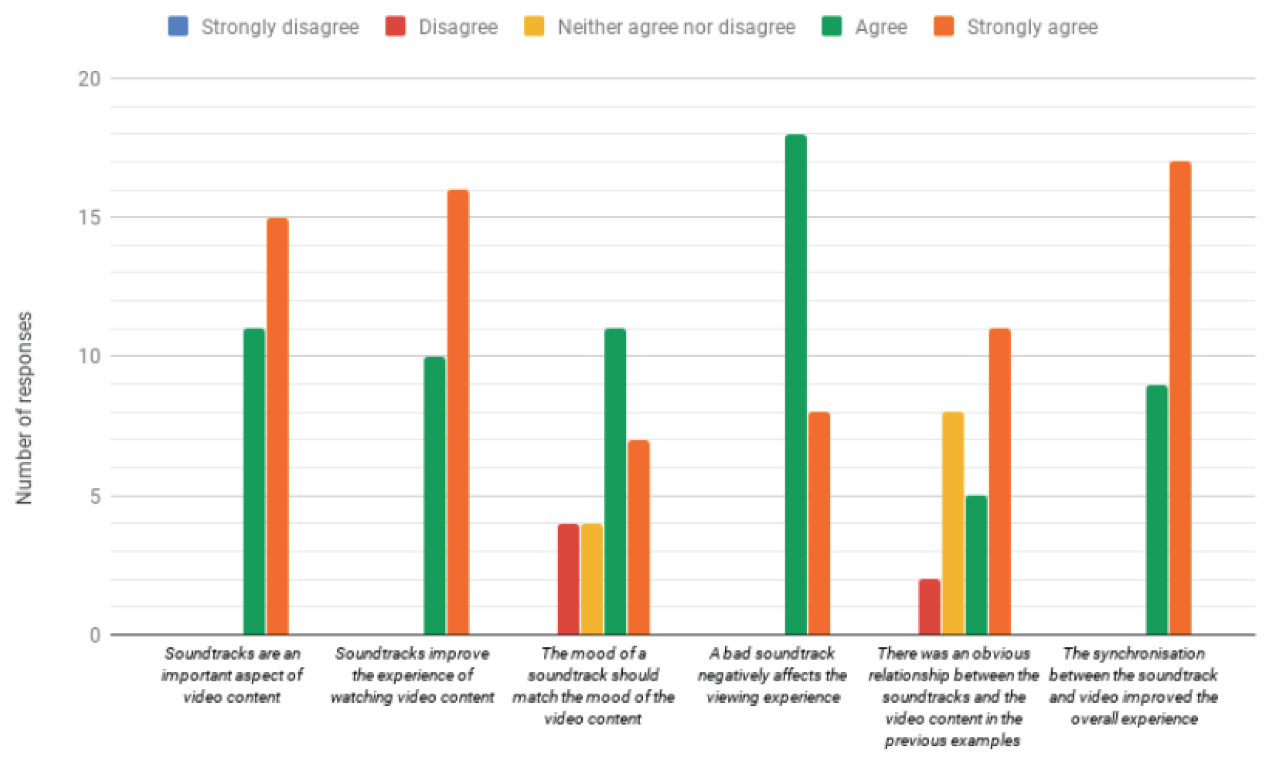}
\caption{Participant responses to all statements in Task 4 of the user study.}
\label{res:q4}
\end{figure}

Figure \ref{res:q4} presents a summary of the responses to the final task of the study that involved indicating a level of agreement with six statements. Generally speaking, participants agreed with all the statements, although some disagreed with the notion that soundtracks should match the mood of the visual content (statement \ref{stm:4-3}) and likewise some did not detect a relationship between the auditory and visual content (statement \ref{stm:4-5}) in the sample videos.

The responses to statement \ref{stm:4-5} were understandable given some participants struggled to notice the temporal synchronisation, while the results from Task 2 suggested the number of objects in the scene was not a strong indicator of mood/intensity.

The key takeaway from this task is that 100\% of participants agree that soundtracks play an important role in experiencing video content. 62\% of participants strongly agreed that synchronisation improves the experience. While these results suggest that Barrington is not yet performing as strongly as hoped for, there is a clear indication that the problems being tackled are relevant to potential users.

A complete breakdown of responses to individual questions in this study can be found in \emph{Anonymous Citation}.

\subsection{Reflections}
\label{sec:reflection}

This user study, while by no means conclusively proving that synchronisation of audio and video provides a better experience than unsynchronised content, was an important step in validating some of the work that has gone into this project. We used a fixed set of questions to disguise which video was the reference and which the impaired. Even when using subjective measures of quality, careful consideration is required to avoid ambiguous questions that could result in different interpretations by different participants. 


Additionally, some interactive tasks could have been used instead of passive questionnaires. One example would be to ask participants to manually mark the points in a video at which they detect a scene transition, in order to objectively evaluate the performance of scene detection algorithms against a human. These interactive tasks could also form the basis of a dataset that could be used to train a machine to tackle some of the challenges with this topic.

This work has shown that temporal synchronisation of audio and video can lead to an enhanced user experience, and that Barrington's synchronisation method was accurate, as perceived by the  participants. Our user evaluation provides some support for hypothesis \emph{H1}. Less successful was the object-counting proxy used to determine the energy of a scene; this method was not able to accurately convey the intended relationship between auditory and visual content to a viewer and we can reject hypothesis \emph{H2}.

The limiting factor of this system is the music generation system used. Watson Beat is the most feature-packed of the open-source options currently available, but a lack of published literature or  documentation makes it a difficult tool to work with and adapt for more general use. As it stands, in terms of musical structure, there is little one can do beyond defining the number and duration of sections within a piece of music. More fine grained temporal control over musical components, such as chord progressions and melodies, would allow for the development of new methods to create a temporal mapping. There is the potential for scene transitions, or even content within a scene, to define how different chords within a piece of music flow. For example, an action shot of a child raising its arms could be used to trigger a musical buildup or the climax of a particular phrase of music, which would more effectively demonstrate audio-visual synchronisation.

Emotionality is an extremely important aspect of music; significant work would be required to enable this system to associate moods and genres of music with visual content. Creating a dataset of images labelled with the same mood and emotion tags as the corpus of music used to train the music generation model would be an excellent first step in being able to associate visual and auditory moods using machine learning. For example, an image classifier trained on such a dataset could potentially identify the overarching mood of a video as happy or sad, using the kind of emotion detection described in \cite{DBLP:journals/corr/abs-1712-02896}. The music generation model would then generate music in a style that has been labelled with the same mood, as opposed to the user-driven approach currently taken in this system.

Barrington is able to automatically generate soundtracks for videos, with the only requirements being a mood and melody on which to base the composition. Scene transition detection and some scene-level object detection form a base structure for the composition to follow. Once generated, a simple post-processing step converts the MIDI representation of the composition into audio, using modifiable instrumentation, after which the soundtrack is ready for use. Processing components can be added or removed as required, and are easy to integrate; for example, we were able to add the object detection functionality in under 40 lines of code. A 30-second soundtrack takes under a minute to generate from an input video, making this a fast, versatile prototype system for testing ideas and a firm foundation for  further work.

\section{Future Work}
\label{sec:concl}



This paper reports on work in progress, and there remain open problems to be tackled in future. 

Adapting and building upon Chu and Roy's work on emotion recognition in movies \cite{DBLP:journals/corr/abs-1712-02896} was something we planned to do during this project, but ultimately was unable to due to the limitations of Watson Beat. Further development of Watson Beat's model of structure is the logical next step, as it would open up opportunities to experiment with new video analysis techniques that can help build a better representation of a video's temporal structure. As musical structure is the limiting factor of a system of this kind, without any advances in capabilities, any further development using current tools would only offer minor improvements.

The results of Task 4 of the user trials (\ref{sec:task4}) confirmed our initial view that a system such as Barrington would be useful to the video content creation community. While the current version is not going to wow viewers with professional quality music, it demonstrates that the ultimate goal of automatically generating emotionally and structurally representative soundtracks from just a video input is valid and achievable.

\bibliographystyle{ACM-Reference-Format}
\bibliography{dassbirdclif}

\end{document}